\documentclass[a4paper,11pt,twocolumn]{article}

\usepackage[utf8]{inputenc}
\usepackage[T1]{fontenc}  

\usepackage{cuted}
\usepackage{graphicx}
\usepackage{amssymb}
\usepackage{mathrsfs}
\usepackage{amsfonts}
\usepackage{amsmath}
\usepackage{textcomp}
\usepackage{epsfig}
\usepackage{subfigure}
\usepackage{endnotes}   
\let\footnote = \endnote
\usepackage{blindtext}
\usepackage{makeidx} 
\usepackage{dblfloatfix}

\usepackage{sidecap}
\usepackage{wrapfig}

\usepackage[table]{xcolor}  
\usepackage{caption}

\usepackage{textcomp}
\usepackage{palatino} 

\usepackage{nomencl}
\usepackage{ifthen}
\usepackage{lipsum}

\makenomenclature

\renewcommand{\nomgroup}[1]
{%
\ifthenelse{\equal{#1}{A}}%
{\item[]\hspace*{-\leftmargin}%
{\textbf{Symboles romains}}}%
{%
\ifthenelse{\equal{#1}{B}}%
{\vspace{3\parsep}\item[]\hspace*{-\leftmargin}%
{\textbf{Symboles grecs}}}%
{%
\ifthenelse{\equal{#1}{C}}%
{\vspace{3\parsep}\item[]\hspace*{-\leftmargin}%
{\textbf{Abréviations}}}%
}
}
}

\usepackage{hyperref}      
\hypersetup{
pdfpagemode=UseOutlines,      
pdfstartview=Fit,             
pdffitwindow=true,            
pdfpagelayout=TwoColumnsRight,
pdftoolbar=true,              
pdfmenubar=true,              
bookmarksopen=false,          
bookmarksnumbered=true,       
colorlinks=true,              
pdfauthor={Ton nom},     
pdftitle={Titre PDF},    
pdfcreator=PDFLaTeX,          %
pdfproducer=PDFLaTeX,         %
linkcolor=blue,               
urlcolor=blue,                
anchorcolor=black,            
citecolor=green,              
frenchlinks=true,             
pdfborder={0 0 0}             
}

\usepackage{multicol}    

\usepackage{amsmath} 
\usepackage{amsthm} 

\usepackage{multirow}
\usepackage{multicol} 
\usepackage[dvips,letterpaper,margin=0.8in,bottom=0.8in]{geometry}

\usepackage{abstract}

\newcommand{\abs}[1]{\left| #1 \right|} 

\let\baraccent=\= 
\renewcommand{\=}[1]{\stackrel{#1}{=}} 

\theoremstyle{definition}

\theoremstyle{remark}

\renewcommand{\thefigure}{\textbf{\arabic{figure}}}
\renewcommand{\figurename}{\textbf{Figure}} 

\usepackage{upgreek}

\usepackage{authblk}

\title{\textbf{Bolometer operating at the threshold for circuit quantum electrodynamics}}

\author[1,$\dagger$]{R. Kokkoniemi }
\author[1,$\dagger$]{J.-P. Girard}
\author[1,3]{D. Hazra}
\author[2]{A. Laitinen}
\author[1,3]{J. Govenius}
\author[1]{R. E. Lake}
\author[1]{I. Sallinen}
\author[1,3]{V. Vesterinen}
\author[2]{P. Hakonen}
\author[1,3,*]{M. Möttönen}

\affil[1]{QCD Labs, QTF Centre of Excellence, Department of Applied Physics, Aalto University, P.O. Box 13500, FIN-00076 Aalto, Finland.}
\affil[2]{Low Temperature Laboratory, QTF Centre of Excellence, Department of Applied Physics, Aalto University School of Science, P.O. Box 15100, FI-00076 Aalto, Finland.}
\affil[3]{VTT Technical Research Centre of Finland Ltd. \& QTF Centre of Excellence, P.O. Box 1000, 02044 VTT, Finland.}
\affil[$\dagger$]{These authors contributed equally.}
\affil[*]{Corresponding author.}

\date{}

\makeatletter
\let\saved@includegraphics\includegraphics
\AtBeginDocument{\let\includegraphics\saved@includegraphics}
\renewenvironment*{figure}{\@float{figure}}{\end@float}
\makeatother

\begin{document}
\twocolumn[
	\begin{@twocolumnfalse}
		\maketitle
\begin{abstract}
\noindent 
Radiation sensors based on the heating effect of the absorbed radiation are typically relatively simple to operate and flexible in terms of the input frequency.
Consequently, they are widely applied, for example, in gas detection\cite{THz_gas_detection_biblio}, security\cite{homeland_security_biblio}, THz imaging\cite{THz_imaging_biblio}, astrophysical observations \cite{astrophysics_J_Wei}, and medical applications\cite{THz_medical_biblio}. 
A new spectrum of important applications is currently emerging from quantum technology and especially from electrical circuits behaving quantum mechanically. This circuit quantum electrodynamics\cite{Blais2020} (cQED) has given rise to unprecedented single-photon detectors\cite{Wallraff_mw_detector_2018,Nakamura_mw_detector_2018}
and a quantum computer supreme to the classical supercomputers in a certain task\cite{Arute2019}. Thermal sensors are appealing in enhancing these devices 
since they are not plagued by quantum noise and are smaller, simpler, and consume about six orders of magnitude less power than the commonly used traveling-wave parametric amplifiers\cite{Macklin_2015}. However, despite great progress in the speed\cite{efetov_nature} and noise levels\cite{Roope_JPA_biblio} of thermal sensors, no bolometer to date has proven fast and sensitive enough to provide advantages in cQED.  
Here, we experimentally demonstrate a bolometer surpassing this threshold with a noise equivalent power of $30\, \rm{zW}/\sqrt{\rm{Hz}}$ on par with the current record\cite{Roope_JPA_biblio} while providing two-orders of magnitude shorter thermal time constant of $500\, \rm{ns}$. Importantly, both of these characteristic numbers have been measured directly from the same device, which implies a faithful estimation of the calorimetric energy resolution of a single 30-GHz photon. 
These improvements stem from the utilization of a graphene monolayer as the active material with extremely low specific heat\cite{specific_heat_graphene}. 
The minimum demonstrated time constant of $200\, \rm{ns}$ falls greatly below the state-of-the-art dephasing times of roughly 100~$\text{\textmu}$s for superconducting qubits\cite{cQEd_coherence_time_100us} 
and meets the timescales of contemporary readout schemes\cite{Walter_2017,Ikonen_2019} thus 
enabling the utilization of thermal detectors in cQED. 
\end{abstract}
\end{@twocolumnfalse}
]


\noindent
Extensive research has been carried out on the development of sensitive THz detectors utilising many different technologies. For example, transition edge sensors\cite{TES_biblio} (TESs), kinetic inductance detectors\cite{KIDs_microwave_biblio} (KIDs), quantum dots\cite{QD_THz_detector}, and qubit-based detectors\cite{Nakamura_mw_detector_2018,Wallraff_mw_detector_2018} have been explored. Especially TESs and KIDs have reached high technological maturity, and are widely applied in astronomy, such as in the observations of the cosmic microwave background\cite{stevens2019_CMB_TES}.
However, detectors for itinerant single microwave photons are still in their early stage of development, mainly due to orders of magnitude lower photon energies requiring higher sensitivity.
Qubit-based detectors have been successfully demonstrated to operate in the single-microwave-photon regime but they have a relatively narrow absorption bandwidth, typically of the order of $10\,\rm{MHz}$, and a dynamic range limited to single photons. In contrast, thermal detectors may provide a large detection bandwidth and dynamic range, and even an energy-resolving detection mode\cite{Pekola2015}.

Advancing thermal detectors towards the single-microwave-photon regime is of great interest to the field of circuit quantum electrodynamics. They could be used, for example, in qubit readout\cite{govia2014high,Opremcak1239} or parity measurement\cite{parity_joonas}. Thermal detectors for qubit readout would be especially beneficial since their readout frequency can be engineered independent of the detection frequency, and consequently they may provide a relief to the frequency crowding challenge in large-scale multiplexing of qubit readout signals; Qubit readout signals even at equal carrier frequencies may be detected with bolometers utilizing frequency multiplexing in their readout.  
In entanglement experiments, photon number eigenstates offer advantages over coherent fields, owing to opportunities in mitigating the effects of loss in the transmission channel\cite{PhysRevX.6.031036,PhysRevLett.114.080503, michael2016new}. 
Such accurate single-shot experiments require a single-photon detector, whereas coherent fields can be detected with linear amplifiers.  Furthermore, a simple thermal detector would greatly decrease the overhead related to the characterization of microwave components\cite{yeh2017microwave, PhysRevX.5.041020, kokkoniemi2017flux} at the single-photon regime. Such characterization is necessary for many components operated at ultralow powers, for example, in quantum computers. 

The sensitivity of radiation detectors is often quantified in terms of noise equivalent power (NEP) which is typically defined as the noise in the readout signal in units of the input power of the detector. Mature technologies, such as TESs and KIDs, have been able to reach NEP in range of a few hundred zW/$\sqrt{\textrm{Hz}}$. We recently introduced a Josephson-junction-based bolometer\cite{joonas_zJ_biblio,Roope_JPA_biblio} exhibiting NEP of 20 zW/$\sqrt{\textrm{Hz}}$ when operated with a nearly quantum-limited amplifier\cite{vesterinen2017lumped}. Furthermore, qubit-based quantum-capacitance detectors\cite{echternach2018single} have recently been reported to have NEP below 10 zW/$\sqrt{\textrm{Hz}}$. Even lower NEP has been expected from semiconducting charge sensors\cite{komiyama2010single}, but full experimental characterization is lacking. Very recently, a calorimeter based on a superconductor--normal-metal--insulator--superconductor junction has been shown to reach the limit of fundamental temperature fluctuations in thermometry, and hence holds great potential for detection of single microwave photons\cite{karimi_reaching_2020}.

A sensitive bolometer typically relies on maximizing the temperature changes induced by absorption of incident photons. To this end, one may minimize the volume of the absorber and fabricate it from a material with low specific heat. In addition, decreasing the thermal conductance from the absorber to its bath increases the low-frequency response at the cost of decreasing the readout speed. Graphene is a two-dimensional (2D) material with unusual thermal properties, which renders it a promising candidate for the realization of a single-microwave-photon bolometer\cite{graphene_review_X_Du}.
Among the wonderful properties of graphene, it has a low electron density of states which leads to a low heat capacity and fast response. At a relatively high temperature of 5 K, extremely fast thermal relaxation time of 35 ps has been reported\cite{efetov_nature} for a graphene-based bolometer. Another recent study on graphene-Josephson-junction-based bolometer\cite{efetov_graphene_JJ_bolometer} carried out at 0.19~K found NEP of 700 zW/$\sqrt{\textrm{Hz}}$ with theoretical thermal time constant down to 0.6 ns. Together these suggest potential for an energy resolution down to a single 32-GHz photon although such an extreme resolution was not measured. In addition, graphene has a low electrical resistance compared with other two-dimensional materials. Importantly, the resistance can be tuned with an electric field, which enables the possibility of precise impedance matching with a planar antenna or a waveguide using, for example, the detector design of refs.~\cite{joonas_zJ_biblio, Roope_JPA_biblio}.

\begin{figure*}[t]
	\centering
	\includegraphics[height=7.2cm]{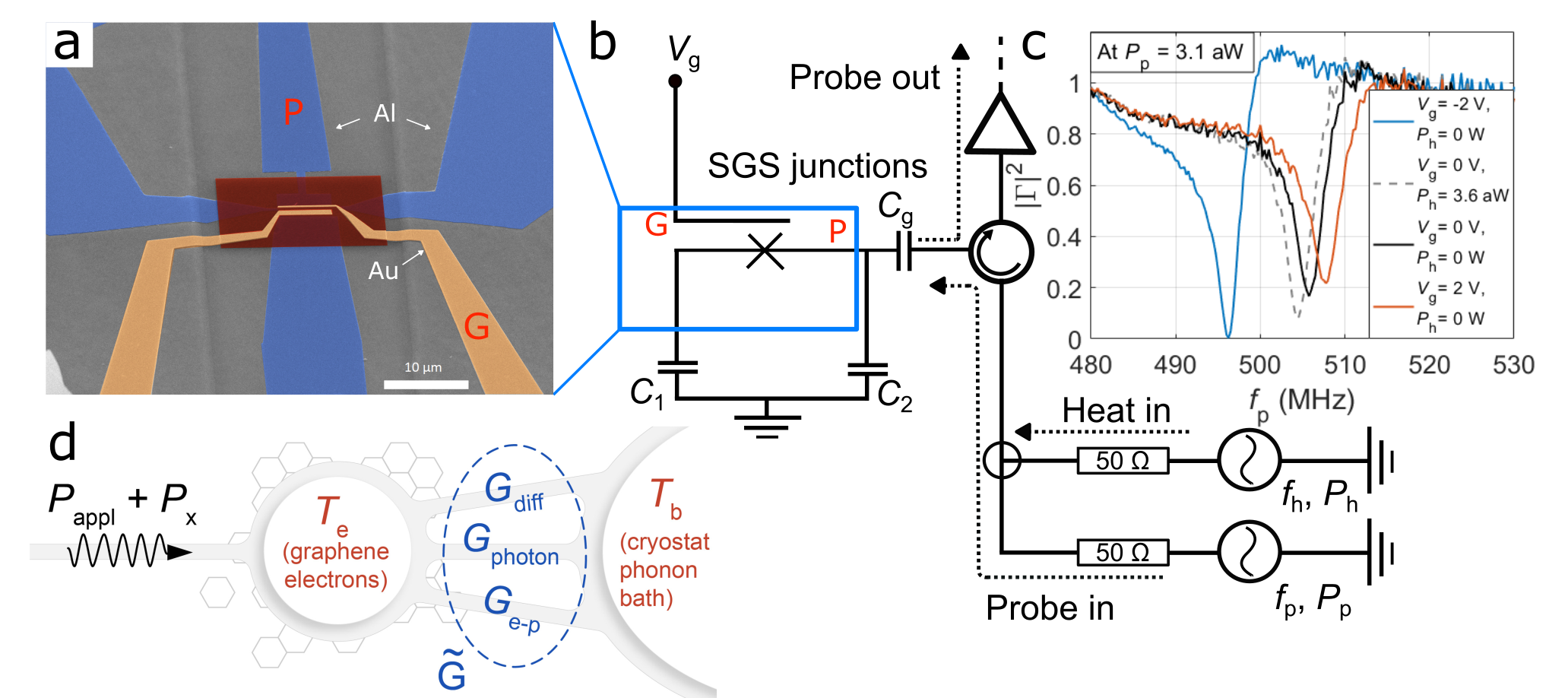}
	\caption{\textbf{Bolometer and its operation principle.} \textbf{a}, False-colour scanning electron microscope (SEM) image of the graphene bolometer. The scale bar denotes $10\, \text{\textmu} \rm{m}$. The gate voltage is applied via port G whereas the heater and probe signals couple through port P to the superconductor--graphene--superconductor (SGS) junction located below the narrowest part of the gate electrode G. Aluminium parts are denoted by blue colour and the gate insulator by red colour. \textbf{b}, Circuit diagram of the detector and a simplified measurement setup. The heater and probe signals, denoted by subscripts h and p, respectively, are combined at room temperature. The microwave reflection coefficient for the probe tone is denoted by $\Gamma$.  \textbf{c}, Reflected fraction of the probe power $P_{\rm{p}}$ as a function of the probe frequency $f_{\rm{p}}$ for the indicated gate voltages $V_{\rm{g}}$ and heater powers $P_{\rm{h}}$ at the bath temperature $T_\textrm{b}=55$~mK. \textbf{d}, Considered thermal model. The electrons in the graphene are coupled to the cryostat phonons through an effective thermal conductance $\tilde{G} = G_{\rm{e-p}}+G_{\rm{diff}}+G_{\rm{photon}}$, which is a sum of the phononic ($G_{\rm{e-p}}$), electron diffusion ($G_{\rm{diff}}$), and photonic ($G_{\rm{photon}}$) thermal conductances (see Methods).}
	\label{scheme_bolometer}
\end{figure*}

In this article, we introduce and demonstrate a hot-electron bolometer based on a superconductor--graphene--superconductor (SGS) junction (Fig.\ref{scheme_bolometer}a). We couple this graphene Josephson junction to on-chip capacitors forming a temperature-dependent $LC$ oscillator (Fig.\ref{scheme_bolometer}b). Incident radiation absorbed in the graphene modifies the resonance frequency of the oscillator, which serves as our thermometer. For example, Fig.\ref{scheme_bolometer}c shows a megahertz-level redshift of the resonance frequency for a heater power of a few attowatts. 

We place a gate electrode on top of the SGS junction allowing us to optimize the charge carrier density in the graphene with electric field. This technique enables us to obtain a low NEP of $30\, \rm{zW}/\sqrt{\rm{Hz}}$ at a thermal time constant of $500\, \rm{ns}$. These results indicate an energy resolution down to a single $30$-GHz photon, which exceeds the performance suggested in ref.~\cite{efetov_graphene_JJ_bolometer}. Importantly, we obtain the NEP and the time constant from direct measurements.
Furthermore, our device exhibits a weak thermal-energy exchange with its environment, with a differential thermal conductance between the graphene bolometer and the phonon bath of the cryostat as low as $0.8\, \rm{fW/K}$, which is less than $2\, \%$ of the quantum of thermal conductance $G_{\rm{Q}}$ at $50\, \rm{mK}$.
The properties of the detector can be tuned with the electric field and the probe signal power and frequency, providing us three degrees of freedom to optimize the performance of the detector. The readout frequency can be tuned by roughly 80 MHz (see Extended Data Fig.~\ref{fig:2D_map_Vg}), and the thermal time constant varies from 200~ns to several micro seconds with probe power and electric field.

\begin{figure}
	\centering
	\includegraphics[height=6.8cm]{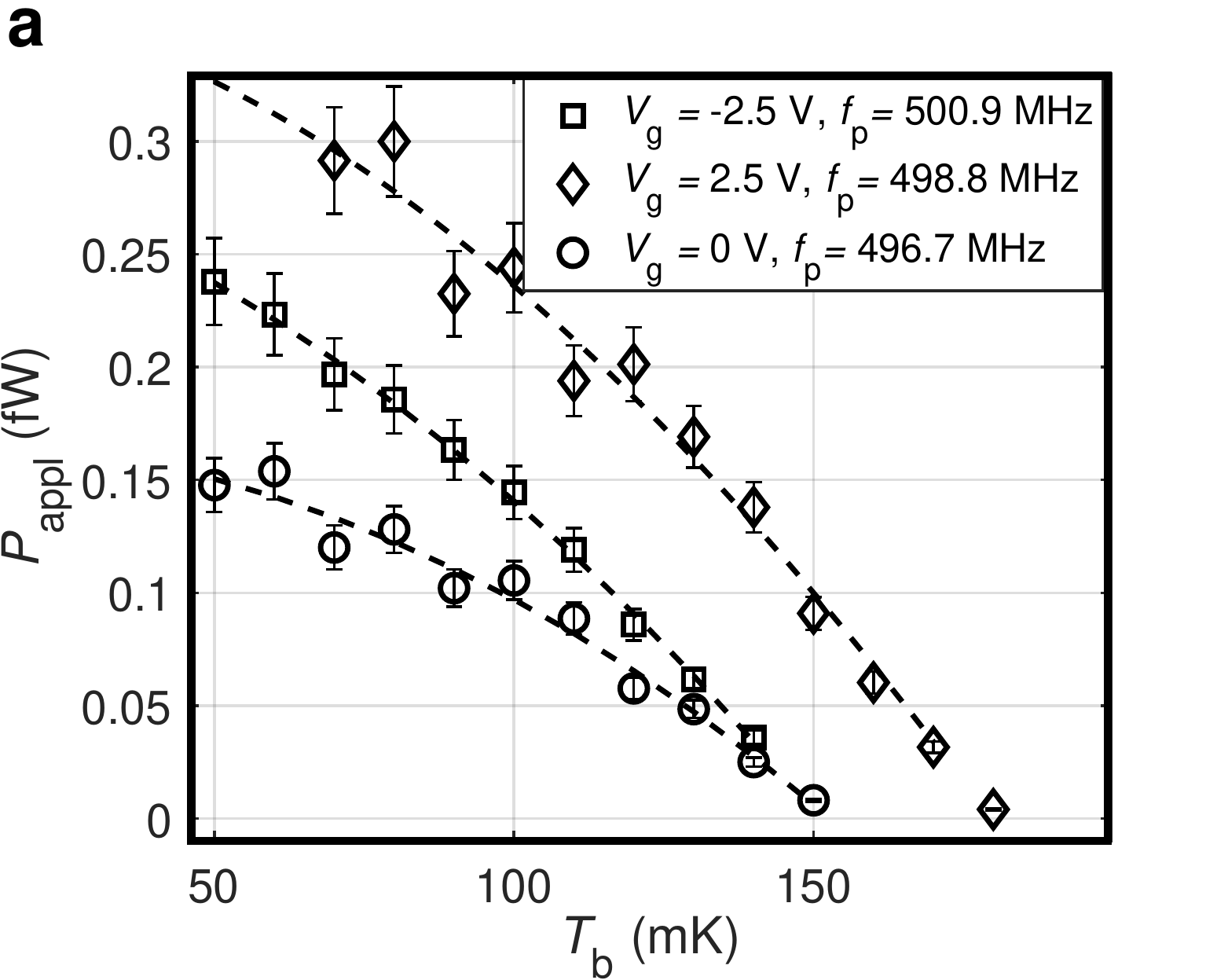}
	\includegraphics[height=6.8cm]{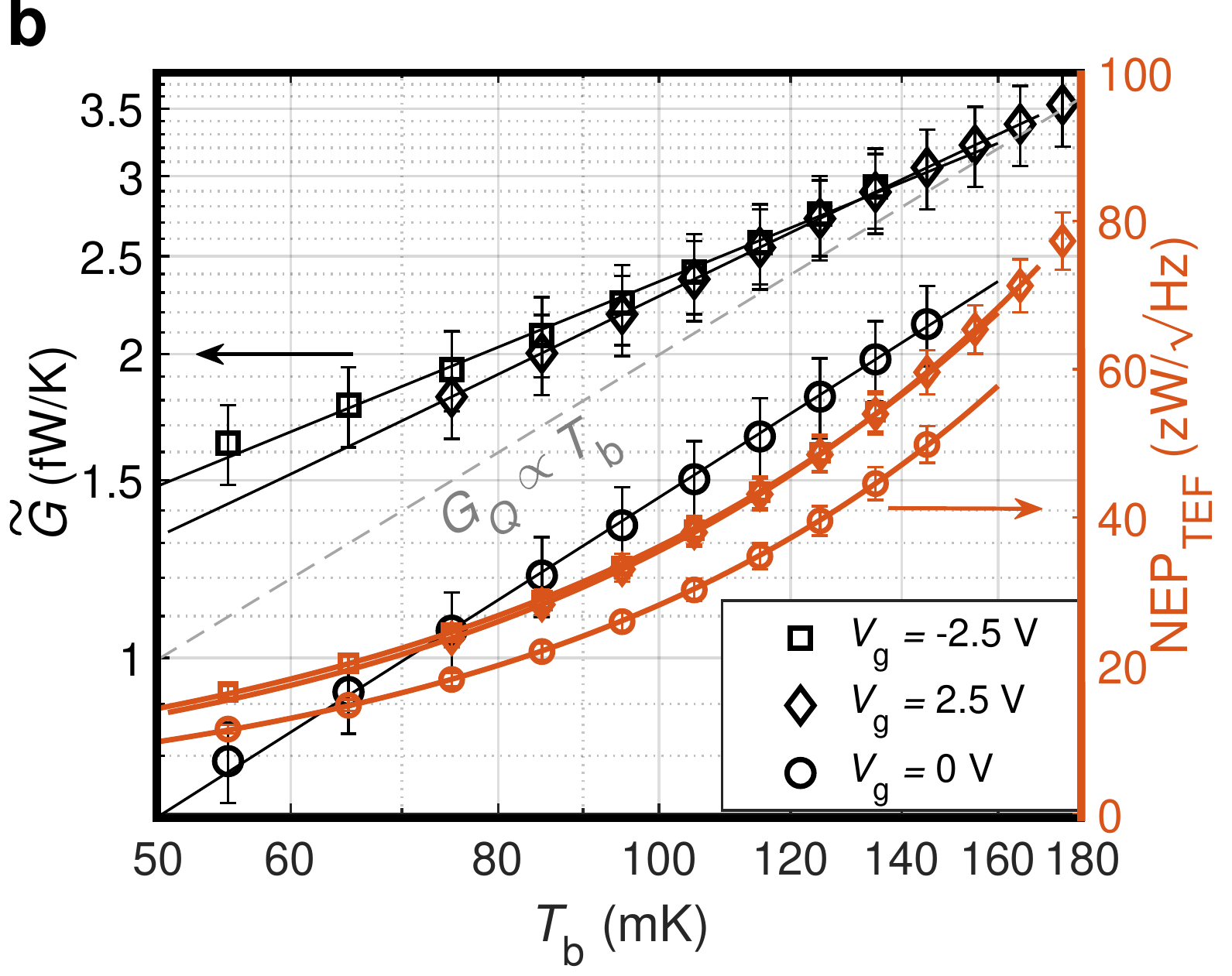}
	\caption{\textbf{Differential thermal conductance and the thermal-fluctuation-limited noise equivalent power.} \textbf{a}, Measured points of constant resonance frequency (markers) in the plane of the cryostat phonon temperature, $T_{\rm{b}}$, and the absorbed power in the graphene, $P_\textrm{appl}$, for the indicated gate voltages and probe frequencies. The dashed lines represent polynomial fits to the data. \textbf{b}, Differential thermal conductance of the graphene electron system (black markers), $\tilde{G}$, as a function of $T_{\rm{b}}$ obtained from the slope of the dashed lines in \textbf{a} at the temperature points of the measured data. The black solid lines are fits to the markers linear on the logarithmic scale. The grey dashed line shows 2.0\% of the quantum of the thermal conductance $G_{\rm{Q}}$. The red markers and lines show the thermal-fluctuation-limited noise equivalent power $\rm{NEP_{TEF}}$ (right vertical axis) corresponding to the differential thermal conductance shown in black colour. The error bars denote $1\sigma$ confidence intervals.}
	\label{thermal_conductance_bolometer}
\end{figure}

Let us discuss in detail our measurements of the differential thermal conductance between the graphene bolometer and the phonon bath of the cryostat.
In a steady state, the heat transfer from the electrons in the graphene at temperature $T_{\rm{e}}$ to the cryostat bath at temperature $T_{\rm{b}}$ is $P_{{\rm{e-b}}} (T_{\rm{e}}, T_{\rm{b}}) = P_{\rm{appl}} + P_{\rm{x}}$, where $P_{\rm{x}}$ is referred to as the parasitic heating, $P_{\rm{appl}} = (1-\abs{\Gamma}^2)P_{\rm{p}}$ is the microwave probe power absorbed by the graphene flake, $\abs{\Gamma}^2$ is the microwave reflection coefficient at the gate capacitor $C_{\rm{g}}$ shown in Fig.\ref{scheme_bolometer}b, and $P_{\rm{p}}$ is the probe power incident on the capacitor. We define the differential thermal conductance by   
$$ \tilde{G} = - \partial_{T_{\rm{b}}} P_{{\rm{e-b}}} (T_{\rm{e}}, T_{\rm{b}}) $$
and measure it by changing the bath temperature and compensating for the resulting change in the resonance frequency, and hence the electron temperature, by changing the applied power as shown in 
Figure \ref{thermal_conductance_bolometer}a. 
Since the electron temperature is constant, the change in the applied power fully flows to the bath, and we obtain the differential thermal conductance in 
Figure \ref{thermal_conductance_bolometer}b as the derivative of the applied power with respect to the bath temperature from Figure \ref{thermal_conductance_bolometer}a.
We observe that $\tilde{G}$ scales at maximum linearly with $T_{\rm{b}}$ as does the quantum of thermal conductance $G_{\rm{Q}} = \pi^2 k_{\rm{B}}^2 T_{\rm{b}}/(3h)$, where $k_{\rm{B}}$ is the Boltzmann constant and $h$ is the Planck constant. This scaling is of significantly lower power in temperature than suggested by studies of electron--phonon coupling in monolayer graphene\cite{phononic_coupling_graphene_Antti,Song_e-phonon_coupling_graphene,Betz_cooling_graphene} which have found $\tilde{G} \propto T_{\rm{b}}^{\delta}$ 
with $\delta \simeq 2-4$ depending on the charge density and the phonon temperature.
This discrepancy tends to indicate that the phononic coupling is not the dominant heat conduction mechanism in our sample. 
The observed behaviour is similarly unlikely to arise from the electron diffusion since the use of superconducting leads to the graphene flake suppresses this effect\cite{Peltonen2010}. Other processes such as multiple Andreev reflections 
may contribute to the heat conduction through the leads\cite{G_diff_with_superconductor}, but their effect is greatly suppressed at the used vanishing voltage bias across the SGS junction. 

However, the observed linear temperature dependence of the thermal conductance may be explained by photonic coupling $G_{\rm{photon}} \propto T_{\rm{b}}$ \cite{electron-photon_conduction_graphene}. This photonic thermal conductance should dominate below the crossover temperature $T_{\text{cr}} = \left[ r_0 \pi^2 k_{\text{B}}^2/(15 h \Sigma S_{\text{graphene}}) \right]^{1/2}$ \cite{electron-photon_conduction_Guichard}, where $r_0$ corresponds to the impedance matching between the detector and its electromagnetic environment, the electron-phonon coupling constant $\Sigma$ is a characteristic of the material, and $S_{\rm{graphene}}$ is the area of the electron gas. Considering a typical value\cite{graphene_Betz} for graphene of $\Sigma = 10^{-15}\, \rm{W \text{\textmu} m^{-2}K^{-4}}$ and an impedance matching $r_0>10^{-2}$, one obtains $T_{\rm{cr}} \geqslant 300\, \rm{mK}$. Thus, by operating at $50\, \rm{mK}<\mathit{T}< 200\, \rm{mK}$ the photonic coupling is likely dominating.

\begin{figure*}[t]
	\centering
	\includegraphics[width=\textwidth]{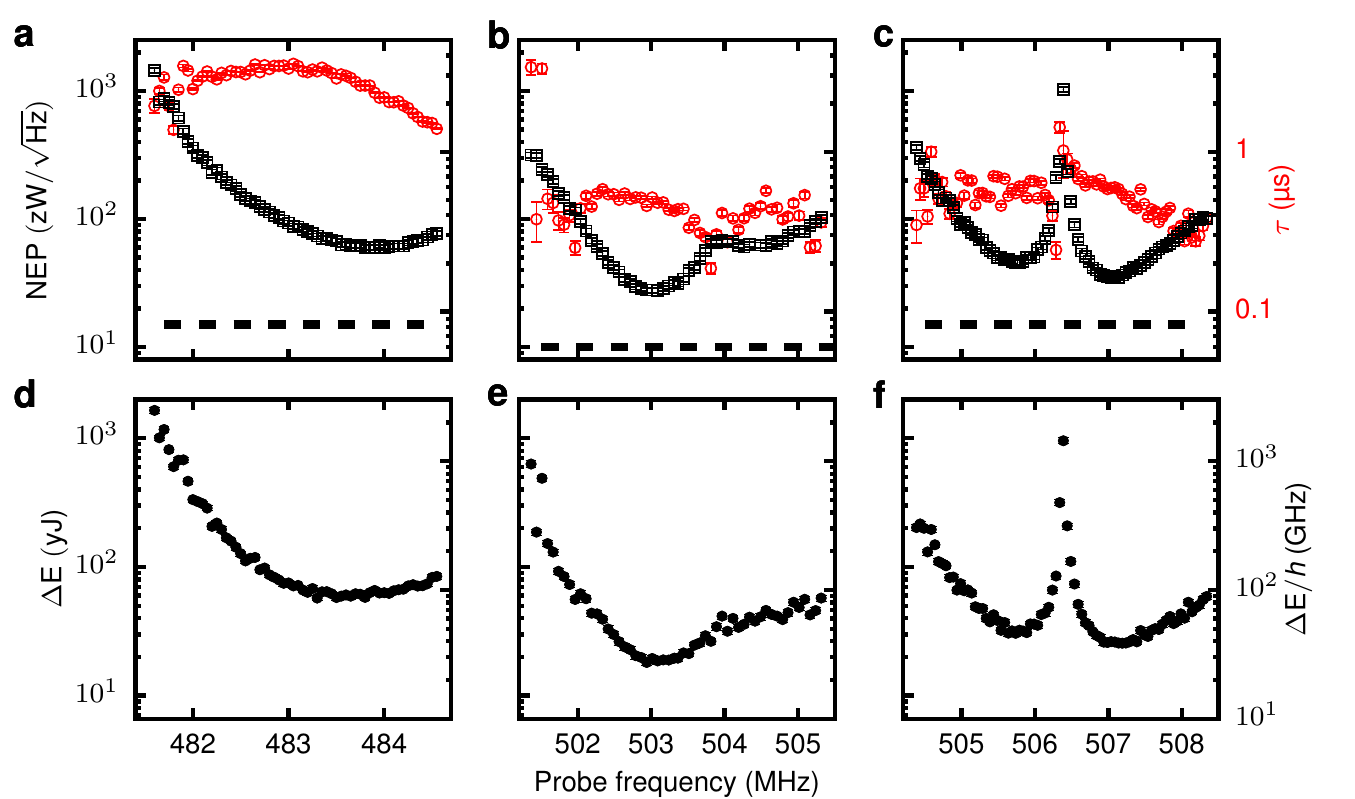}
	\caption{\textbf{Key characteristic properties of the bolometer.} \textbf{a--c}, Measured noise equivalent power (NEP) (black markers, left axis) and thermal relaxation time $\tau$ (red markers, right axis) of the detector for gate voltages, $V_{\rm{g}} = -2.5\, \rm{V}$ (\textbf{a}), $V_{\rm{g}} = 0\, \rm{V}$ (\textbf{b}), and $V_{\rm{g}} = 2.5\, \rm{V}$ (\textbf{c}), bath temperature $T_\textrm{b}= 55$~mK, and probe power $P_\textrm{p}=530$~aW (\textbf{a}), $P_\textrm{p}=140$~aW (\textbf{b}),  and $P_\textrm{p}=370$~aW (\textbf{c}). The dashed horizontal lines indicate the thermal-fluctuation-limited NEPs obtained from the thermal conductance given by Fig.~\ref{thermal_conductance_bolometer}b. The error bars denote $1\sigma$ confidence intervals. \textbf{d-f}, Energy resolution of the bolometer obtained from the NEP and time constant experiments of panels \textbf{a}--\textbf{c}, respectively. See text for details.}
	\label{NEP}
\end{figure*}

We define the NEP to be the noise density in the readout signal in units of the absorbed power. The NEP is obtained in practice as the voltage noise in the readout signal divided by the voltage responsivity of the detector to the absorbed power. For convenience, we only measure the quasistatic responsivity and divide it by $\sqrt{1+\left(2\pi \tau f_{\rm{n}}\right)^2}$ which takes into account the thermal cut-off in the responsivity for noise frequencies $f_{\rm{n}}$ higher than inverse thermal relaxation time $1/\tau$. This is a generally accepted method for obtaining the responsivity, justified by our observations of exponential thermal relaxation dominated by a single time constant (see Extended Data Fig.~\ref{fig:example_time_trace}). 

Figures~\ref{NEP}a--\ref{NEP}c show the experimentally obtained NEP and time constant as functions of the probe frequency at three different gate voltages.
The minimum NEP occurs at a gate voltage of $V_{\rm{g}} = 0\, \rm{V}$ for a probe frequency of $f_{\rm{p}}=503\, \rm{MHz}$ and equals $\rm{NEP} = 30\, \rm{zW}/\sqrt{\rm{Hz}}$. Fortunately, this low NEP coincides with an exceptionally short thermal time constant, $500\, \rm{ns}$. The minimum observed time constant is 200~ns. Note that the thermal time constant yields the speed, at which the bolometer exchanges energy with its environment, but does not pose a fundamental limit on the operation speed of the device in detecting energy packets. Namely, if the internal thermalization of the electrons in the bolometer is fast, the rising edge of the readout signal can be orders of magnitude faster that the falling edge set by the thermal time constant. 
Thus the measured thermal time constant seems promising for applications  in cQED with the state-of-the-art readout time of the order of 100~ns. 

Random exchange of energy quanta with the environment of the bolometer leads to fluctuations in the local electron temperature, and hence may significantly contribute to the total noise in bolometers\cite{Penttil__Phonon_noise,karimi_reaching_2020}. 
The thermal-fluctuation-limited NEP is given by\cite{doi:10.1063/1.357128} $\rm{NEP}_{\rm{TEF}} = \sqrt{4 \mathit{k}_{\rm{B}}\mathit{T}_{\rm{e}}^2 \mathit{\tilde{G}} }$. Using $T_{\rm{e}} = 55\, \rm{mK}$,  $V_{\rm{g}}=0\, \rm{V}$, and the data of Fig.~\ref{thermal_conductance_bolometer}b, we obtain $\rm{NEP}_{\rm{TEF}} = 12\, \rm{zW}/\sqrt{Hz}$. This indicates that our bolometer operates close to the thermal bound. Thus improvements of roughly a factor of two in the minimum NEP may be obtained by technical changes of the measurement setup such as was obtained in ref.~\cite{Roope_JPA_biblio} by the introduction of a nearly quantum-limited amplifier. However, major further progress calls for development of the device itself such as redesign of the sample or use of new materials. For example, reduction of the photonic heat conduction by advanced filtering schemes is likely to improve the NEP.

In order to maximise the absorption coefficient, we operate the heater signal at a frequency close to the resonance frequency, typically around $500\, \rm{MHz}$. On the other hand, this optimization prevents us from measuring directly the single-photon calorimetric energy resolution of the device which would require much higher energies for the absorbed photons. However, we extract the energy resolution using the NEP\cite{Moseley1984Thermal, enss2005cryogenic}, $\Delta E = \left( \int_0^{\infty} \frac{4 \textrm{d}f_n}{\text{NEP}(f_n)^2}  \right)^{-1/2}$, and show the results in Figs.~\ref{NEP}d--\ref{NEP}f. The finest energy resolution of $20\, \rm{yJ}$ is obtained at $55\, \rm{mK}$. This corresponds to the energy of a single $30$-GHz photon, or alternatively five 6-GHz photons which is a satisfactory frequency and photon number scale for the usual readout of superconducting qubits.
The bolometer seems also an ideal detector candidate in an alternative qubit readout scheme demonstrated in ref.~\cite{Opremcak1239}, where the total energy subject to the first-stage measurement device may be even higher than in the usual qubit readout, for example, $100\textrm{ yJ}=h\times 150\textrm{ GHz}$ in ref.~\cite{Opremcak1239}. 

Alternatively, a lower bound of the energy resolution can be estimated from the thermal conductance and time constant. A simple thermal model with a single thermal relaxation time, as shown in Fig.\ref{scheme_bolometer}d, is enough to describe our observations. Thus we may estimate the heat capacity of the graphene as $C_{\rm{e}} = \tilde{G} \tau$. For $f_{\rm{p}}=503\, \rm{MHz}$, $V_{\rm{g}}=0\, \rm{V}$, $\tau=500\, \rm{ns}$ and $T_{\rm{b}} = 50\, \rm{mK}$, we obtain $C_{\rm{e}} = 2.5 \times 10^{-22} \rm{JK}^{-1}$, which corresponds to $1.2\times 10^{-23} \, \rm{JK}^{-1}\text{\textmu}m^{-2} =0.87\times k_\textrm{B}\text{\textmu}m^{-2}$. It is in a rather good agreement with the literature value of roughly $2\mathit{k}_{\rm{B}}\, \text{\textmu} m^{-2}$ at $50\, \rm{mK}$ assuming a linear temperature dependence of the heat capacity\cite{specific_heat_graphene}. From this heat capacity we estimate the standard deviation of energy owing to the fundamental thermal fluctuations using the equation $\Delta E_\textrm{th} = \sqrt{\mathit{k}_{\rm{B}} \mathit{T}_{\text{e}}^{2} \mathit{C}_{\rm{e}} } = 3\, \rm{yJ}= \mathit{h} \times 4.4\, \rm{GHz}$ at $50\, \rm{mK}$. 


This article experimentally demonstrates an ultrafast low-noise graphene bolometer operating in the microwave range: measured thermal conductance as low as $0.8\, \rm{fW/K}$, thermal relaxation time $\tau$ down to $200\, \rm{ns}$, and NEP as low as $30\, \rm{zW}/\sqrt{\rm{Hz}}$ which is close to the corresponding thermal-fluctuation limit $\rm{NEP}_{\rm{TEF}} = 12\, \rm{zW}/\sqrt{\rm{Hz}}$.  
The achieved low noise level and fast response surpass the threshold for applications in circuit quantum electrodynamics where ultralow powers are used and detected in time scales orders of magnitude shorter than the typical state-of-the-art coherence times of roughly 100~{\textmu}s\cite{cQEd_coherence_time_100us}.  
Our experiments indicate an energy resolution of our device in the yoctojoule range in the calorimeter mode. It sets the detection threshold at a single $30$-GHz microwave-photon level which is, to the best of our knowledge, the finest resolution reported for a thermal detector. 
Interestingly, the estimated low heat capacity of the bolometer implies a thermal energy uncertainty of only $h\times 4.4$~GHz which suggest that such a fine energy resolution 
may be achievable already with the present device by technical improvements in the measurement setup such as integration of a nearly quantum-limited amplifier to the readout circuit\cite{Roope_JPA_biblio}. In the future, we aim to use bolometers in the framework of circuit quantum electrodynamics to study quantum phenomena with photon-number-based detection, free of quantum noise stemming from the Heisenberg uncertainty relations. Furthermore, unusual Josephson physics\cite{Wiedenmann2016} seems interesting for utilisation in bolometers.


\bibliographystyle{naturemag}
\bibliography{bib_ref}{}

\section*{Methods}

\textbf{Sample fabrication} \\
First, a commercial 4” Si substrate with resistivity exceeding $10\, \rm{k} \Omega$cm is covered with a 300-nm-thick thermally grown silicon oxide ($\rm{SiO_x}$). Then a 200-nm-thick film of niobium (Nb) is deposited onto the substrate by dc magnetron sputtering. Microwave resonators are fabricated using standard photolithography and reactive ion etching of the niobium film. Subsequently, 50-nm-thick aluminium oxide ($\rm{AlO_x}$) is grown by atomic layer deposition, covering the entire wafer. The parallel-plate capacitors are patterned and fabricated using standard electron beam lithography, aluminium (Al) evaporation and lift-off. Here, the $\rm{AlO_x}$ layer serves as the insulator, separating the Nb ground plane from the top Al electrode fabricated simultaneously with the Al capacitor plates.

After cleaving the chips from the wafer, the graphene flake is placed on the wafer using a micromanipulator technique. Then the Ti/Al leads from the graphene to the capacitor plates are defined using e-beam evaporator using poly(methyl methacrylate) (PMMA) mask. The Cr/Au electrode, which is used to gate the carrier density in the graphene, is separated from the flake by a $150\text{-}\rm{nm}$ PMMA layer which behaves as an insulator.\\ 

\noindent \textbf{Sample and measurement setup}\\
The experiments are carried out in a commercial cryostat operating at a temperature of $55\, \rm{mK}$. The heater and the probe signal generators are connected at room temperature to a single cable which channels the corresponding microwaves to the bolometer through several filters and attenuators. In our scheme, the function of the heater signal is to heat the SGS junction by photon absorption whereas the probe signal is mostly reflected off the device, amplified and digitized. A part of the heater signal is also reflected because of an unavoidable impedance mismatch in our setup. To avoid contaminating the readout signal with the reflected heater signal, we deliberately desynchronize the clock of the heater signal generator from the clocks of the probe signal generator and digitizer. Therefore, the heater signal is averaged out from the readout signal.

Figure \ref{scheme_bolometer}a shows a coloured SEM image of the nanobolometer. The central element of the bolometer is the graphene flake (total surface area of order $21\, \text{\textmu} \rm{m}^2$) with the SGS junction located below the gate electrode labelled G and the PMMA layer (red). The SGS junction is connected to a $50\text{-}\Omega$ aluminium superconducting transmission line through a port labelled P. The SGS junction is capacitively coupled to the ground and from circuit modelling we estimate its intrinsic inductance to be $2.32\, \rm{nH}$, for $T_{\rm{b}}= 55\, \rm{mK}$ at low probe power, 0 V gate bias, and without heater power. We estimated the value of the capacitances $C_1$, $C_{2}$, and $C_{\rm{g}}$ (see Extended Data Fig.~\ref{photo_sample_capacitors}) from the design parameters to be $124\, \rm{pF}$, $57\, \rm{pF}$, and $1.4\, \rm{pF}$, respectively.
We estimate the charge carrier density in graphene to be $n = 1.3 \times 10^{12}\, \rm{cm^{-2}} $. This value is deduced from Fig \ref{fig:2D_map_Vg}a since the Dirac point is not reached for $\abs{V_{\rm{g}}}<20\, \rm{V}$, which corresponds to $n = C_{\rm{i}} V_{\rm{g}}/e$ with $C_{\rm{i}}\approx 2.36\times 10^{-4}\, \rm{Fm^{-2}}$ the capacitance between the graphene flake and the gold electrode. 

From Fig.\ref{scheme_bolometer}c, we measure the probe transmission coefficient $S_{21}$ to determine $\Gamma$. To normalize the data, we measured $S_{21}$ with the gate voltage set to $V_{\rm{g}}= -12\, \rm{V}$ and with high probe and heater power in order to move the resonance frequency of the bolometer and its tank circuit outside the frequency range considered. Note that increasing the detector temperature is also an alternative way to move out the resonance peak.
Extended Data Fig.~\ref{fig:2D_map_Vg} presents the reflection coefficient $\abs{S_{21}}$ of the probe signal, as function of the gate voltage and the probe power. We observe that the resonance frequency can be tuned by the gate voltage by roughly 80 MHz. Similar behaviour occurs with the probe power, where the redshift is caused by an absorption of a fraction of the probe signal which results in an increase of $T_{\rm{e}}$. \\

\noindent \textbf{Heater power calibration} \\
In order to calibrate the attenuation of the heater input line shown in Extended Data Fig.~\ref{fig:full_measurement_setup}, we have placed microwave switches inside the cryostat which allow us to effectively replace the sample with a 50-$\Omega$ resistor or a direct short to ground. 
By replacing the sample with a $50\text{-}\Omega$ resistor, we measure the gain of the output amplification chain up to the room temperature mixer using the so-called Y-factor method: 
We connect a spectrum analyzer (Rohde \& Schwarz FSV 40) to this point in the chain, and measure the noise from the $50\text{-}\Omega$ resistor as a function of its temperature. Fitting a straight line to the measured power as a function of temperature yields the gain and the noise temperature of the amplification chain. For these measurements, we set the spectrum analyzer resolution bandwidth to 10 MHz and took 10000 averages.
Next, we replace the sample with a direct short to ground and measure the transmission amplitude from the heater signal generator up to the mixer using a vector network analyzer with resolution bandwidth set to 10 Hz and averaging to 21.
Substracting the gain of the amplification chain from the transmission measurement yields the attenuation of the heater input line as shown in Extended Data Fig.~\ref{fig:heater_calibration}

In addition to the attenuation in the heater line, the absorbed heater power is further decreased since a part of the signal is reflected from the bolometer. To account for this effect, we measure a reference transmission coefficient from the heater signal generator to the digitizer with high heating power and such a gate voltage that the resonance frequency is far away from the studied frequency range, and hence the bolometer is essentially fully reflecting the heater tone. In addition, we measure the transmission coefficient for every data point in Fig.~\ref{NEP}. The ratio between these data and the reference transmission coefficient yields the reflection coefficient of the bolometer for the heater tone, $\Gamma_\textrm{h}$, from which we obtain the absorbed fraction of the heater power reaching the bolometer as $1-|\Gamma_\textrm{h}|^2$. The absorbed heater power is obtained by multiplying the heater power applied at the generator by this fraction and by the measured attenuation of the heater line. The absorbed heater power is used in all measurements of the responsivity of the detector. This is justified by the fact that it is straightforward to engineer essentially perfectly impedance-matched heater line to this type of a bolometer as shown in refs.~\cite{joonas_zJ_biblio,Roope_JPA_biblio}. We leave this engineering task for future work.\\

\noindent\textbf{Differential thermal conductance} \\
The differential thermal conductance of an electron system at temperature~$T_\textrm{e}$ is defined as the partial derivative $\partial_{T_{\rm{e}}} P_{{\rm{e-b}}}$ of the heat flowing out of the electron system into its bath at temperature $T_{{\rm{b}}}$. Here, the derivative it taken with respect to the electron temperature such that the bath temperature is constant.
Due to the small size of the graphene flake however, the measurement of $T_{\rm{e}}$ in our system has a large uncertainty, for example, due to parasitic heat contributions which may dominate at temperatures below 50 mK. Thus instead, we compute the differential thermal conductance from  
\[ \tilde{G} = - \partial_{T_{\rm{b}}} P_{{\rm{e-b}}} (T_{\rm{e}}, T_{\rm{b}}) \]
where the electron temperature is considered constant.
This is valid as long as we consider linear channels for the flow of the heat currents \cite{joonas_zJ_biblio,Joonas_bolometer}, with no rectification. 

In Fig. \ref{scheme_bolometer}(d), we show the thermal model for heat transfer between $T_{\rm{e}}$ and $T_{\rm{b}}$. The differential thermal conductance $\tilde{G}$ is a sum of three contributions: the electron-phonon coupling $G_{\rm{e-p}}$, the electron diffusion into the superconducting leads $G_{\rm{diff}}$, and the electromagnetic coupling with the environment $G_{\rm{photon}}$. For simplicity, we consider the accurately measurable cryostat bath temperature, $T_\textrm{b}$, to coincide with the temperatures of all baths directly coupled to the electron system of the graphene flake.
Thus the thermal effects of the system are effectively described by a single heat capacity $C_{\rm{e}}$ directly coupled to the cryostat phonon bath with a thermal conductance $\tilde{G}$. Our observations in Extended Data Fig.~\ref{fig:example_time_trace} of a single thermal time constant $\tau$ support the validity of this simple thermal model.

The method we use to obtain $\tilde{G}$ is based on mapping contours of a constant monotonically $T_{\rm{e}}$-dependent quantity $A(T_{\rm{e}})$ in the $T_{\rm{b}}-P_{\rm{appl}}$ plane and taking a derivative of the contour\cite{Joonas_bolometer}, thus satisfying the above condition that the electron temperature must be constant. Here, we choose the resonance frequency as the $T_{\rm{e}}$-dependant quantity since is decreases monotonically with increasing electron temperature. 

In practice, we measure the contours of constant resonance frequency by searching for the heater power required to keep the resonance frequency constant given a change in the bath temperature (Figure \ref{thermal_conductance_bolometer}a). This is a slow measurement because we need to change the temperature of the whole mixing chamber plate of the cryostat with several kilograms of metal. Consequently, the measurement is affected by significant $1/f$ noise. Although the effect of this noise is reasonable in the contours of constant electron temperature in Figure \ref{thermal_conductance_bolometer}a, its relative contribution would become too large if we would take a simple numerical derivative of the contours to obtain the differential thermal conductance. Thus instead, we make a polynomial fit to the data in Figure \ref{thermal_conductance_bolometer}a and differentiate the fitted function. In Figure \ref{thermal_conductance_bolometer}b we show these results along with power law fits to the differential thermal conductances motivated by such typical behaviour in the heat conduction channels.\\

\noindent \textbf{Extraction of the regime of linear response}\\
Our definition of NEP assumes that the response of the detector is linear with respect to the heater power, i.e., the responsivity of the detector is constant. Therefore, we measure the response as a function of heater power at the ($P_\mathrm{p}, f_\mathrm{p}$) point where we find the lowest NEP for each gate voltage. The results are shown in Extended Data Fig.~\ref{fig:linearity}. We find linear response up to 16 aW with -2.5 V gate bias, and 4 aW with 0 V and 2.5 V bias. \\

\noindent \textbf{Data acquisition and processing} \\
After reflection from the bolometer, amplification, and filtering, the 500-MHz probe signal is converted down to 70.3125 MHz by the analog electronics shown in Extended Data Fig.~\ref{fig:full_measurement_setup}. This intermediate-frequency signal is digitized at a rate of $250\times 10^6$~samples per second by NI-5782 connected to NI-7972R field-programmable gate array (FPGA). The recorded samples are digitally down converted  to dc on the FPGA. Due to technical limitations, our FPGA runs at clock frequency of 125 MHz, and therefore we do the down conversion to two samples in parallel. Consequently, we obtain both the in-phase (I) and quadrature (Q) signals for both the probe and the reference channel, resulting in total eight channels of data. Subsequently, we combine the two parallel samples owing to slow FPGA clock to bring the total number of data channels down to four. We also have the option to combine temporally adjacent samples on the FPGA. Typically we combine in total four samples, thus effective digitization rate is 62.5 MHz. The FPGA collects a pre-determined number of samples and also carries out ensemble-averaging of all four data channels. For data presented in Fig.~\ref{NEP}, we took 262144 ensemble-averages for the time traces, and noise spectrum was averaged over 30 repetitions. For the linearity measurements presented in Extended Data Fig.~\ref{fig:linearity}, we took 32768 averages.

After the averaging, the data is transferred to a desktop computer. We rotate the I and Q signals such that the bolometer response appears completely in the in-phase signal~I. For technical reasons, we further apply a digital low-pass filter to the data with a cut-off frequency of 2~MHz. 
Extended Data Fig.~\ref{fig:example_time_trace} 
shows the filtered signal in a typical experiment where we apply a heater pulse for a certain period of time to the bolometer to measure the thermal time constant and dc response. 
To this end, we fit an exponential function to the filtered data separately for the rising and the falling edge since the time constant of the edges can, in general, be different due to electrothermal feedback. In the NEP and energy resolution calculation, we use the time constant obtained for the rising edge.
For the presented data in Fig.\ref{thermal_conductance_bolometer} of the differential thermal conductance, we took 16384 averages.

\section*{Acknowledgements}
We acknowledge the provision of facilities and technical support by Aalto University at OtaNano – Micronova Nanofabrication Center and LTL infrastructure which is part of European Microkelvin Platform (EMP, No. 824109 EU Horizon 2020).
We have received funding from the European Research Council under Consolidator Grant No. 681311 (QUESS) and under Advanced Grant No. 670743 (QuDeT), European Commission through H2020 program projects QMiCS (grant agreement 820505, Quantum Flagship), the Academy of Finland through its Centers of Excellence Program (project Nos. 312300, 312059, and 312295) and grants (Nos. 314447, 314448, 314449, 276528, 305237, and 314302), the Finnish Cultural Foundation, and the Vilho, Yrjö and Kalle Väisälä Foundation of the Finnish Academy of Science and Letters. We thank Wei Liu, Matti Partanen, and Leif Gr\"onberg for assistance in nanofabrication and useful discussions. 



\section*{Author contributions}
R. K. and J.-P. G. conducted the experiments and data analysis. The sample was designed by R. E. L. and fabricated by D. H. and A. L. Initial characterizations were carried out by I. S., D.H, and J. G. Majority of the measurement code was written by V. V. and J. G. The manuscript was written by R. K., J.-P. G., and M. M. with the help of comments from all authors. The work was actively supervised by P. H. and M. M. 

\renewcommand{\figurename}{\textbf{Extended Data Fig.}} 
\setcounter{figure}{0} 

\clearpage

\begin{figure*}
    \centering
    \includegraphics[height=6cm]{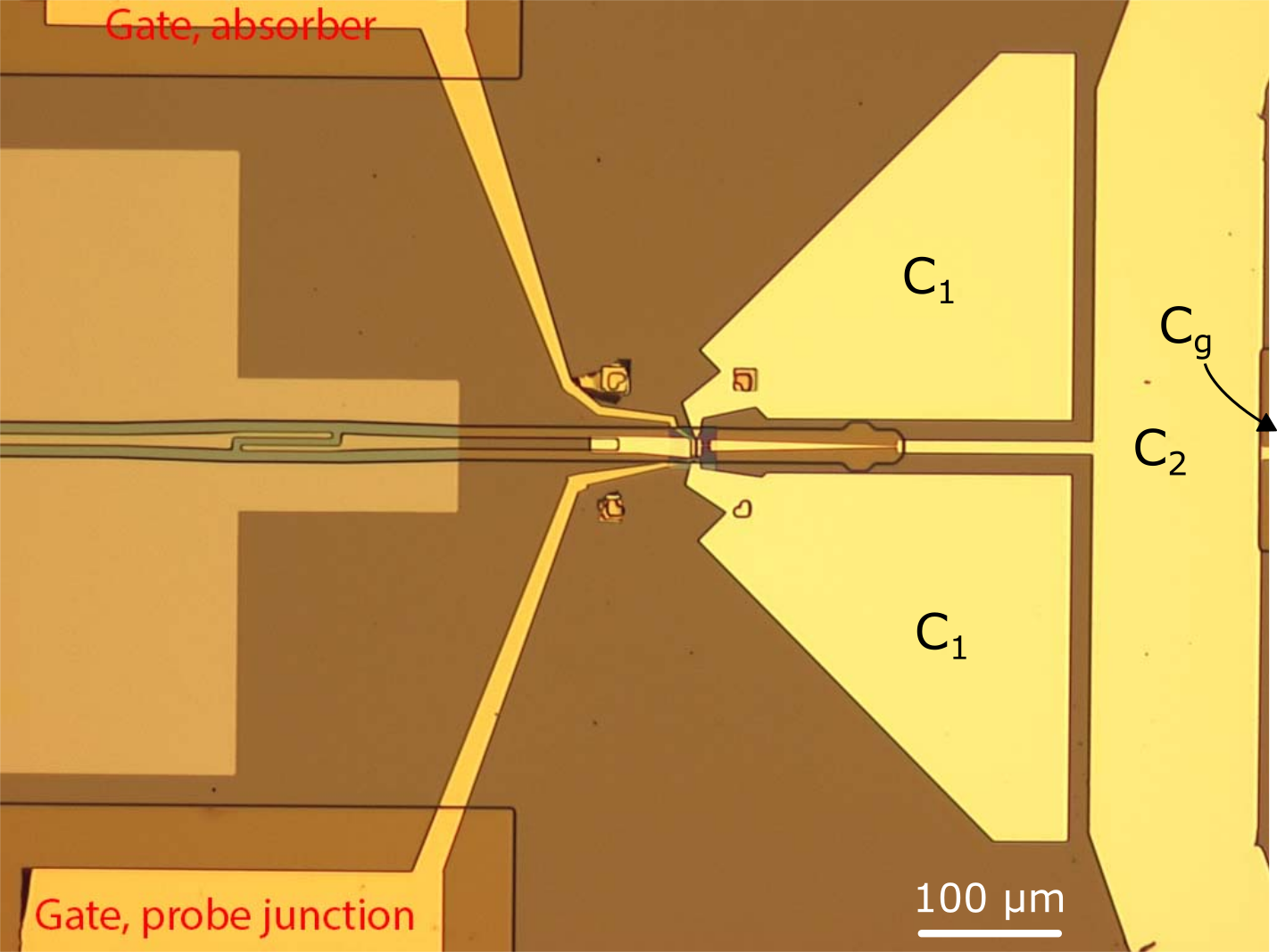}
    \caption{\textbf{On-chip details of the bolometer.} Optical image of the bolometer indicating the locations of the capacitors $C_1$, $C_2$, and $C_\textrm{g}$. The transmission line on the centre left and the absorber gate line on the top left were not used in the experiment.}
    \label{photo_sample_capacitors}
\end{figure*}

\clearpage

\begin{figure*}
    \centering
    \includegraphics[width=8.8cm]{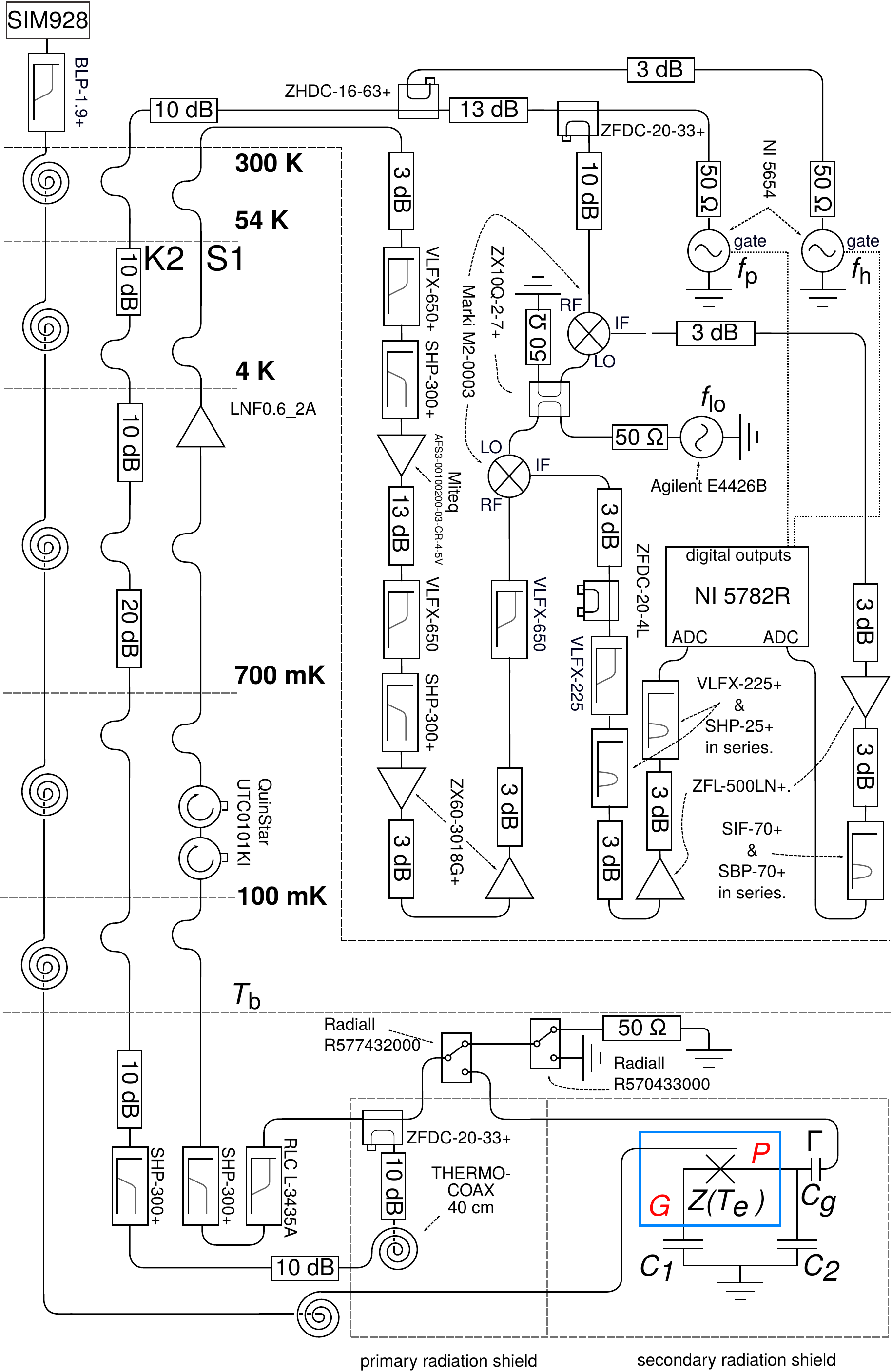}
    \caption{\textbf{Detailed measurement setup.} The blue rectangle corresponds to the SEM image displayed in Fig.~\ref{scheme_bolometer}a. The model numbers on the filters, couplers, and  amplifiers refer to Mini-Circuits product numbers. Nominal temperatures of the different cryostat plates (dashed horizontal lines) are indicated. The 180-degree bends and spirals refer to long segments in the co-axial microwave cables and Thermocoax cables, respectively, which function as thermal bottlenecks separating the different temperature stages. The top-most dashed line refers to the boundary of the cryostat. The nested millikelvin radiation shields are made of gold-plated copper.}
    \label{fig:full_measurement_setup}
\end{figure*}

\clearpage

\begin{figure*}
    \centering
    \includegraphics[width=8.8cm]{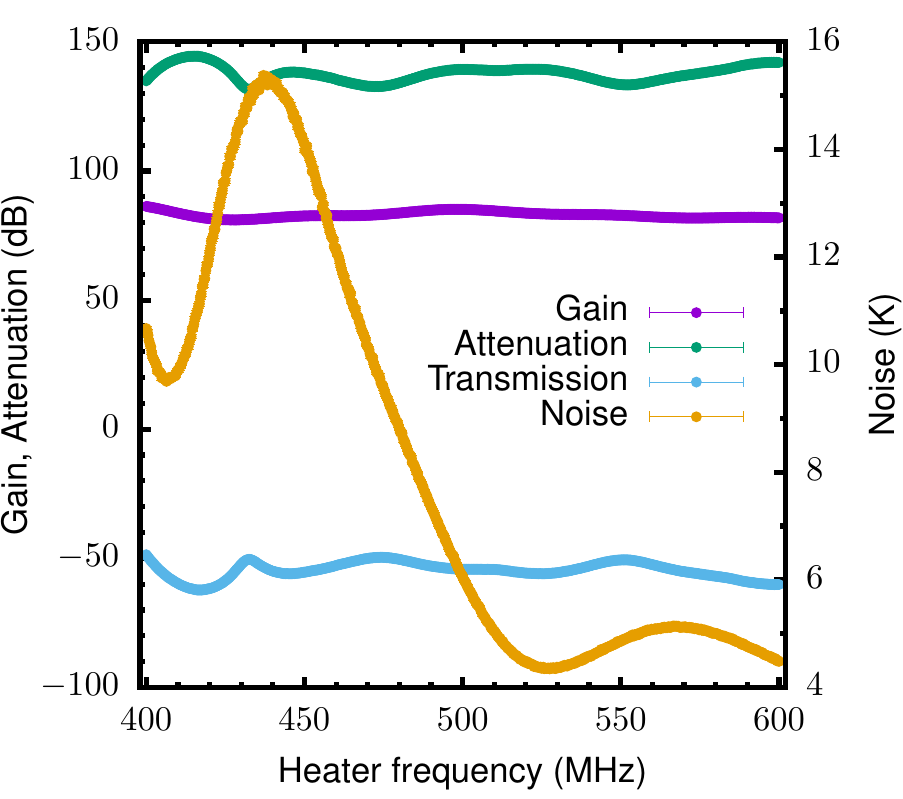}
    \caption{\textbf{Calibration of the heater power.} Gain (left axis) and noise temperature  (right axis) of the amplification chain from the microwave switch shown in Extended Data Fig.~\ref{fig:full_measurement_setup} to the room temperature mixer, transmission amplitude from the heater signal generator to the mixer  (left axis), and attenuation of the heater line from the signal generator to the switch (left axis) as functions of heater frequency. The error bars denote $1\sigma$ confidence intervals.}
    \label{fig:heater_calibration}
\end{figure*}

\clearpage

\begin{figure*}[t]
    \centering
    \includegraphics[width=8.7cm]{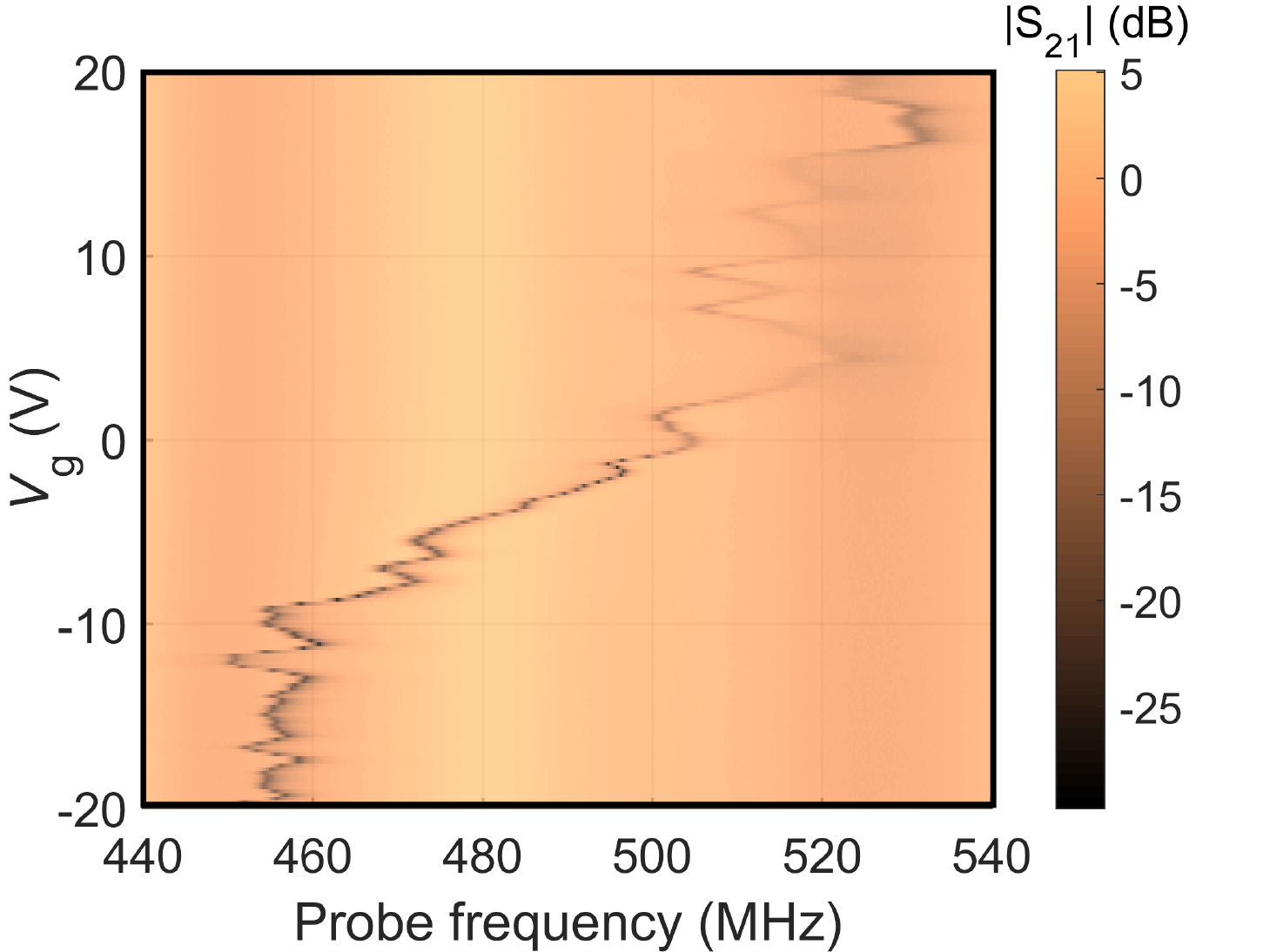}
    \includegraphics[width=8.7cm]{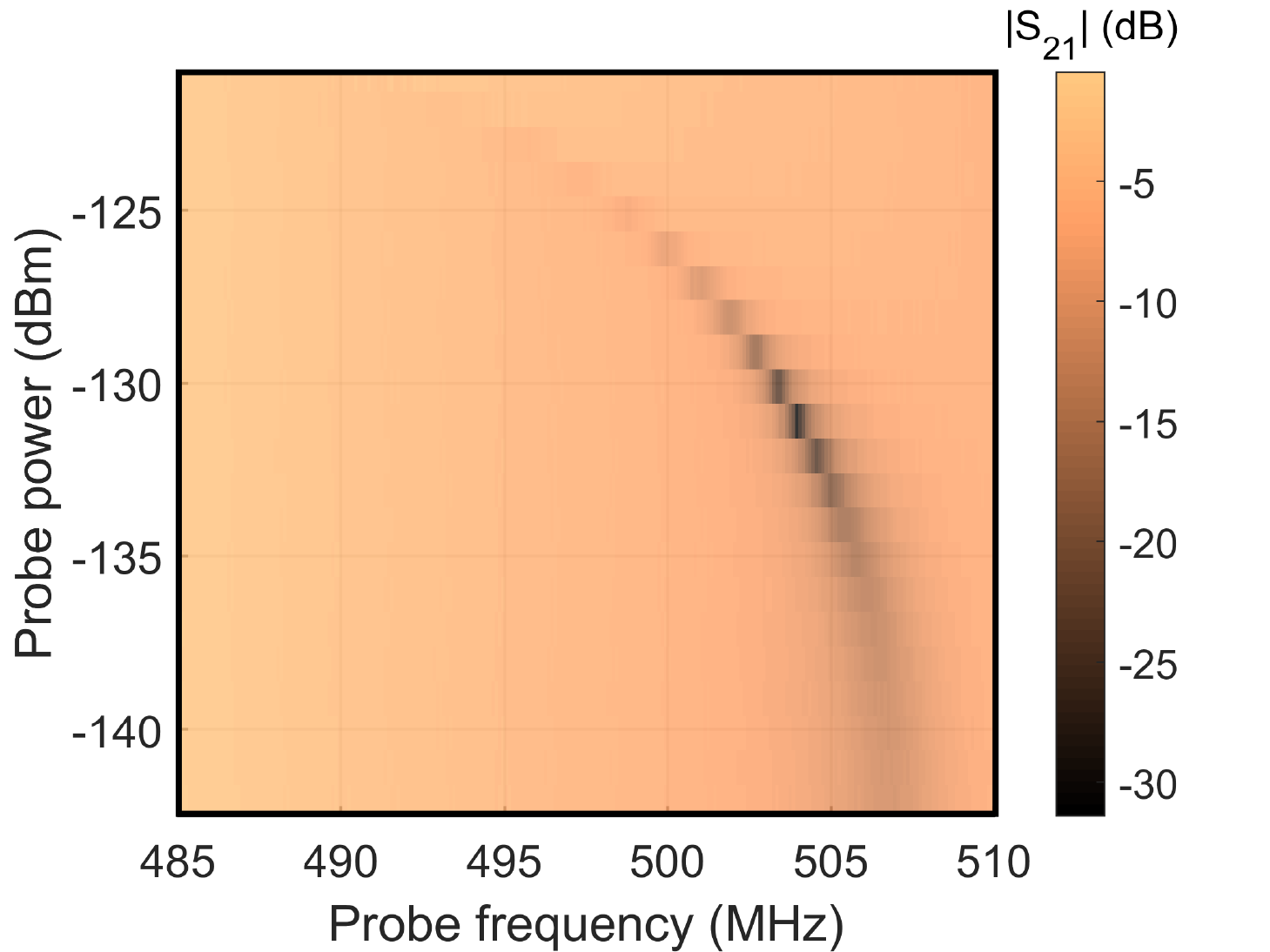}
    \caption{\textbf{Controlled shift of the resonance frequency.} \textbf{a}, Measured reflection amplitude of the bolometer as a function of the probe frequency and the gate voltage at $P_{\rm{p}}= 17.4\, \rm{aW}$. \textbf{b}, Reflection amplitude as a function of the probe frequency and power at $V_{\rm{g}}= 0\, \rm{V}$. In each panel and each horizontal trace, the point of the lowest reflection amplitude roughly yields the resonance frequency. The bath temperature is $T_\textrm{b}=55$~mK and no heater power is applied.}
    \label{fig:2D_map_Vg}
\end{figure*}

\clearpage

\begin{figure*}[h]
    \centering
    \includegraphics[width=17cm]{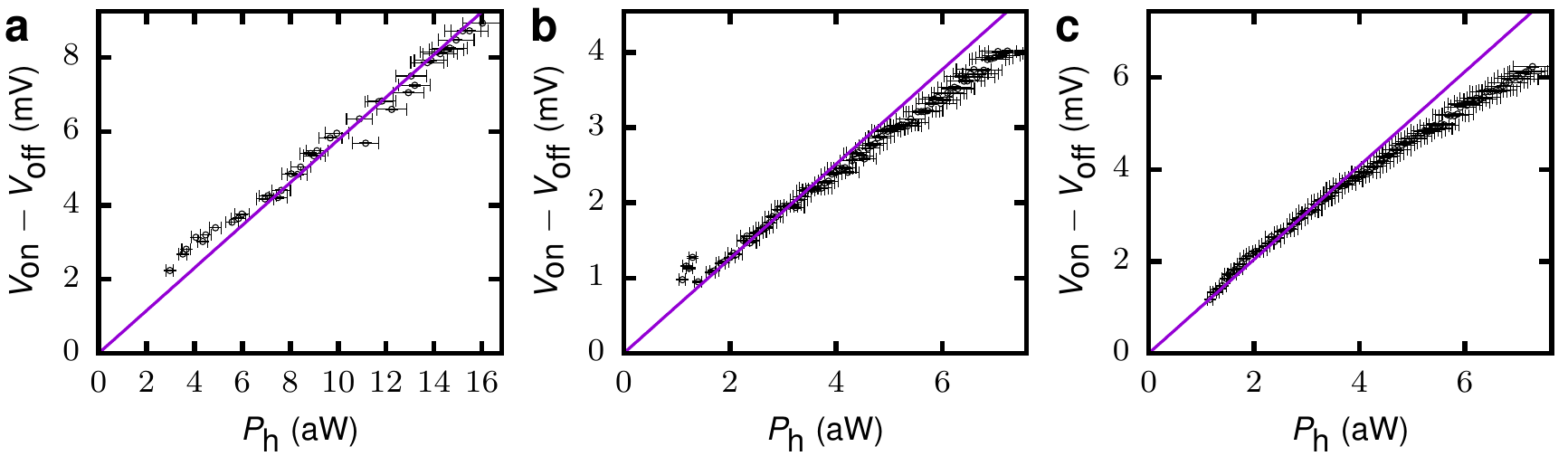}
    \caption{\textbf{Response to heater power.} \textbf{a--c,} Difference in the voltage of the measured probe signal quadrature between heater on and heater off as a function of the heater power level in the on state for gate voltage -2.5 V (\textbf{a}), 0 V (\textbf{b}), and 2.5 V (\textbf{c}) at the probe power and frequency where we find the lowest noise equivalent power in Figure~\ref{NEP}. The solid line shows a linear fit up to heating power of 16 aW (\textbf{a}) and 4 aW (\textbf{b}) and (\textbf{c}). The error bars denote $1\sigma$ confidence intervals.}
    \label{fig:linearity}
\end{figure*}

\clearpage

\begin{figure*}
    \centering
    \includegraphics[width=7cm]{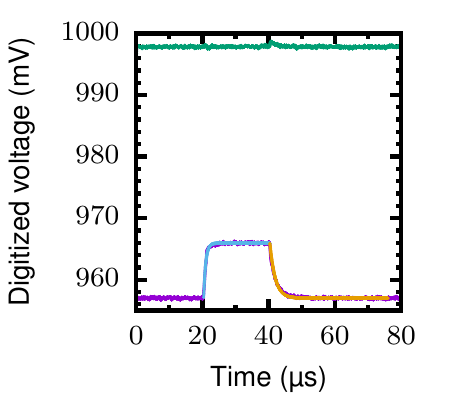}
    \caption{\textbf{Extraction of the thermal time constant and voltage response.} 
    Example time trace of the real (purple markers) and imaginary (green markers) parts of the measured probe signal for the heater power quickly turned on at the time instant $t=20$ \textmu s and turned off at $t=40$ \textmu s. An exponential fit to the rising edge of the real part is denoted with cyan colour and a fit to the falling edge with orange.}
    \label{fig:example_time_trace}
\end{figure*}

\clearpage
\section*{Supplemental material}

\renewcommand{\thefigure}{\textbf{S\arabic{figure}}}
\renewcommand{\figurename}{\textbf{Supplementary Fig.}} 
\setcounter{figure}{0}

\noindent \textbf{Circuit model} \\
In order to obtain more insight on the characteristic properties of the superconductor-graphene-superconductor (SGS) junction, 
we used a simplified circuit model where the SGS junction is represented by a parallel connection of an inductor $L$ and a resistor $R$, the impedance of which is $ Z = \left[ (i\omega L)^{-1}+R^{-1} \right]^{-1}$. Consequently, the circuit load impedance $Z_{\rm{L}}$ is calculated by taking into consideration the capacitances $C_1$, $C_2$, and $C_{\rm{g}}$ as
\[ Z_{\rm{L}} = (i\omega C_{\rm{g}})^{-1} + \left\{ i\omega C_2 + \left[ (i\omega C_1)^{-1} + Z) \right]^{-1} \right\}^{-1} \]
This model reproduces the position and the depth of the resonance but does not account for the asymmetry observed in the measurement. A better agreement with the experimental data would require to consider the electrothermal feedback effect \cite{joonas_zJ_biblio}. Nevertheless, this model provides a relatively good estimate on the parallel resistance and inductance of the SGS junction as shown in Supplementary Fig.~\ref{circuit_modeling}. The characteristic inductance and resistance are 2~nH and 10 k$\Omega$, respectively.
\newpage
\begin{figure}[h!]
	\centering
	\includegraphics[height=6cm]{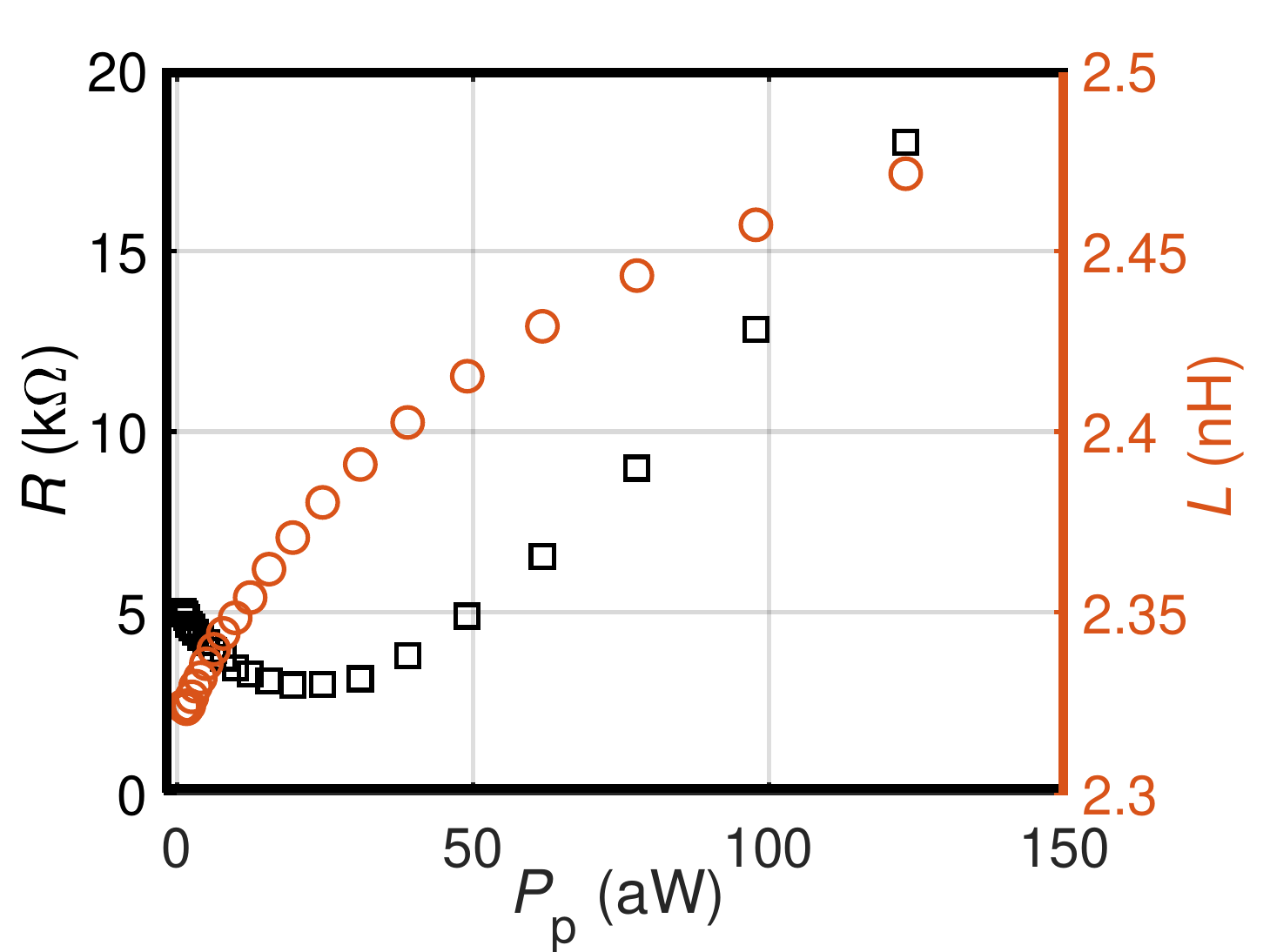}
	\caption{\textbf{Electrical-equivalent parameters of the superconductor-graphene-superconductor (SGS) junction.} Effective parallel resistance (black rectangles, left axis) and inductance (orange circles, right axis) of the SGS junction as functions of the probe power $P_{\rm{p}}$, at gate voltage $V_{\rm{g}}=0\, \rm{V}$, heater power $P_\textrm{h}=0$~aW, and bath temperature $T_{\rm{b}}=50\, \rm{mK}$. These values are extracted from experimental microwave reflection data such as that shown in Figure~\ref{scheme_bolometer}c by fitting the reflection coefficient from a linear circuit model to the measured results (see text).}
	\label{circuit_modeling}
\end{figure}

\end{document}